\documentclass[usegraphicx,usenatbib,useapjfonts,apj,twocolappendix]{emulateapj}
\usepackage{graphicx}
\usepackage{amsfonts}
\usepackage{pifont}

\def\alp2{\mbox{$\alpha_{\rm 0.02}$}}
\def\cmtres{\mbox{cm$^{-3}$}}
\def\esf{\mbox{{$\epsilon_{\rm SF}$}}}
\def\erg{\mbox{erg}}

\def\Ke{\mbox{K}}

\def\lcdm{\mbox{$\Lambda$CDM}}
\def\lwdm{\mbox{$\Lambda$WDM}}

\def\mh{\mbox{$M_{\rm v}$}}
 
\def\mpch{\mbox{$h^{-1}$Mpc}} 
\def\ms{\mbox{$M_{s}$}}
 
\def\msunh{\mbox{$h^{-1}$M$_\odot$}} 
\def\msun{\mbox{M$_\odot$}}

\def\nsf{\mbox{$n_{\rm SF}$}}
\def\ome{\mbox{$\Omega_m$}} 
\def\omeb{\mbox{$\Omega_b$}}
\def\omel{\mbox{$\Omega_\Lambda$}}

\def\re{\mbox{$R_{e}$}}
\def\r2{\mbox{$r_{-2}$}}
\def\rh{\mbox{$R_{\rm v}$}}

\def\ro2{\mbox{$\rho_{\rm -2}$}}

\def\con2{\mbox{$C_{\rm -2}$}}
\def\cnfw{\mbox{$C_{\rm NFW}$}}
\def\sige{\mbox{$\sigma_8$}} 
\def\Tsf{\mbox{{$T_{\rm SF}$}}}
\def\vmax{\mbox{$V_{\rm max}$}}

\def\mwdm{\mbox{$m_{\rm WDM}$}}

\def\mhm{\mbox{$M_{f}$}}
\def\wdma{\mbox{WDM$_{1.2}$}}

\def\mathnew{\mathsurround=0pt} 
\def\simov#1#2{\lower .5pt\vbox{\baselineskip0pt 
    \lineskip-.5pt\ialign{$\mathnew#1\hfil##\hfil$\crcr#2\crcr\sim\crcr}}}   
\def\simgreat{\mathrel{\mathpalette\simov >}}  
\def\simless{\mathrel{\mathpalette\simov <}}  
\def\'#1{\ifx#1i{\accent"13\i}\else{\accent"13#1}\fi}

\defcitealias{alej2014}{G+2014}

\def\plotdelgado#1{\includegraphics[width=8.9cm]{#1}}
\def\plotdelgados#1{\includegraphics[width=8.7cm, height=9.0cm]{#1}}

\shorttitle{The inner structure of dwarf sized halos}
\shortauthors{Gonzalez-Samaniego et al.}  

\begin{document}

\title{The inner structure of dwarf sized halos in Warm and Cold Dark Matter cosmologies} 

\author{ A. Gonz\'alez-Samaniego\altaffilmark{1,2},V. Avila-Reese\altaffilmark{1}, P. Col\'{\i}n\altaffilmark{3}}

\altaffiltext{1}{Instituto de Astronom\'{\i}a, Universidad Nacional Aut\'onoma de M\'exico, 
A.P. 70-264, 04510, M\'exico, D.F., M\'exico}

\altaffiltext{2}{Present address: Center for Cosmology, Department of Physics and Astronomy, University of California, Irvine, CA 92697, USA}

\altaffiltext{3}{Instituto de Radioastronom\'{\i}a y Astrof\'{\i}sica, Universidad Nacional 
Aut\'onoma de M\'exico, A.P. 72-3 (Xangari), Morelia, Michoac\'an 58089, M\'exico }

\begin{abstract}
By means of N-body+Hydrodynamics zoom-in simulations we study the evolution of the
inner dark matter and stellar mass distributions of central dwarf galaxies formed in halos of virial masses 
$\mh=2-3\times 10^{10}$ \msunh\ at $z=0$, both in a Warm Dark Matter (WDM) and Cold Dark Matter (CDM)
cosmology. The half-mode mass in the WDM power spectrum of our simulations is $\mhm= 2\times 10^{10}$ \msunh. 
In the dark matter (DM) only simulations halo density profiles are well described by the NFW parametric fit in 
both cosmologies, though the WDM halos have concentrations lower by factors of 1.5--2.0 than their
CDM counterparts. In the hydrodynamic simulations, the effects of baryons significantly flatten the inner density,
velocity dispersion, and pseudo phase-space density profiles of the WDM halos but not of the CDM ones. 
 The density slope, measured at $\approx 0.02$\rh, \alp2, becomes shallow in periods of 2 to 5 Gyr in the WDM 
runs. We explore whether this flattening process correlates with the global star formation (SF), \ms/\mh\ ratio, gas outflow, and
internal specific angular momentum histories. We do not find any clear trends, but when \alp2\ is shallower
than $-0.5$, \ms/\mh\ is always between 0.25 \% and 1 \%. We conclude that the main reason
of the formation of the shallow core is the presence of strong gas mass fluctuations inside the inner halo, 
which are a consequence of the feedback driven by a very bursty and sustained SF history in shallow gravitational potentials. Our WDM halos, which
assemble late and are less concentrated than the CDM ones, obey these conditions. There are also
(rare) CDM systems with extended mass assembly histories that obey these conditions and
form shallow cores. The dynamical heating and expansion processes behind the DM core flattening
apply also to the stars in a such a way that the stellar age and metallicity gradients of the dwarfs are softened, 
their stellar half-mass radii strongly grow with time, and their central surface densities decrease. 

\end{abstract}


\section{Introduction}
\label{intro}

Determining the nature of dark matter is one of the biggest challenges in astrophysics, cosmology
and particle physics. From the point of view of cosmic structure formation, the Cold Dark Matter proposal
(the so-called \lcdm\ cosmology) is very successful at large scales, but at the scales of low-mass galaxies, 
apparent disagreements with some observational inferences have been reported. Thus, variations on 
the nature of dark matter have been discussed to alleviate the potential issues with this theory
\citep[for a review, see e.g.,][and the references therein; hereafter Paper I]{Colin+2015}. 
However, the predictions at small scales are controversial given 
the complex physics of baryons (e.g., star formation and its feedback) and the strong interplay between dark 
and baryonic matter
at these scales. For example, several authors have proposed that the effects of 
baryons in low-mass \lcdm\ halos with strong supernova (SN)-driven outflows or with bursty star formation (SF) events, which drive 
periodical bulk gas motions in the inner $\sim 1$ kpc region, are able to significantly flatten the cuspy halo 
mass density profiles \citep[e.g.,][]{Navarro+1996,Read+2005,Mashchenko+2006,Pontzen+2012}. 
Shallow halo cores have been proposed to be produced also due to the transfer of
energy and angular momentum to the inner dark matter by subhalos or baryonic clumps falling to the center via dynamical friction 
\citep[e.g.,][]{ElZant+2001,Romano-Diaz+2008,delPopolo+2009,Inoue+2011,Cole+2011,delPopolo+2014,Nipoti+2015}.

In modifying the cosmological model, the introduction of Warm Dark Matter (WDM) instead of 
CDM has been extensively studied since the first N-body cosmological simulations 
were presented 
within this scenario \citep[][]{Moore+1999,Colin+2000,Avila-Reese+2001,Bode+2001}. The halos/subhalos obtained in 
the N-body simulations do not show any evidence of shallow cores, although they are overall less concentrated 
than their CDM counterparts if the mass of the halo is not much greater than the filtering mass 
of the WDM power spectrum \citep[][]{Avila-Reese+2001, Colin+2008, Lovell+2012, Schneider+2012,Anderhalden+2013}.

Whether or not significant shallow halo cores are produced in \lcdm\ cosmological N-body + hydrodynamics (hereafter hydro) 
simulations has been a matter of intense debate in the last years. Given that the subgrid physics in these 
simulations is not well constrained, the results partially depend on the choice of the SF, feedback
schemes, and parameters, as well as on the initial conditions of the zoom-in simulations. 
The formation or not of shallow cores has been reported to depend on several factors (likely related to each other)
such as the amplitude and fluctuation level of the SF rate (SFR),
the stellar-to-halo mass ratio, variations of 
the baryonic or gas mass in the innermost halo regions, etc. 
\citep[e.g.,][]{Mashchenko+2008,Governato+2010,Governato+2012,Maccio+2012,Teyssier+2013,Madau+2014, 
diCintio+2014,Onorbe+2015,Tollet+2015,Chan+2015}. 
In contrast to the hydro simulations on which these works
are based, there are others where a significant flattening of the inner halo 
mass distribution has not been observed or still others that require too much SN feedback efficiency
in order to attain it \citep{Garrison-Kimmel+2013,Vogelsberger+2014,Gonzalez-Samaniego+2014,Schaller+2015}.

In Paper I, we have presented zoom-in hydro simulations of dwarf galaxies formed in 
WDM halos of masses close to the filtering scale, \mhm,\footnote{The filtering mass is defined as 
$\mhm = (4 \pi /3) \bar\rho (\lambda_{\rm hm}/2)^3$, where $\lambda_{\rm hm}=2\pi/k_{\rm hm}$ and 
$k_{\rm hm}$ are the comoving half-mode length and wavenumber, that is where the value of the power 
spectrum of the WDM model is half that of the corresponding CDM one; $\bar\rho$ is the present-day 
background density.} 
as well as masses $20-30$ \mhm. While in the latter case the structure formation 
proceeds as in the CDM scenario, in the former, the evolution and properties of galaxies 
yield very different results in the two scenarios. 

In the present paper, we discuss the evolution of the halo/galaxy mass density and velocity dispersion
profiles of the WDM and CDM simulations presented in Paper I. In particular, we study 
the question of shallow core formation in both scenarios. 
Our results show that for distinct halos of present-day virial masses $2-4\times 10^{10}$ \msun, 
shallow cores form in the WDM scenario (with \mhm\ around these masses) while in the corresponding
CDM simulations this is not the case. \citet{Herpich+2014} and \citet{Governato+2015} have also presented 
simulations of WDM and CDM galaxies. These authors concluded that galaxy properties and evolution are more sensitive to stellar feedback than they are to WDM candidate mass. In the case of the dwarf galaxy simulated in \citet{Governato+2015}, a shallow core forms in the halo for both the CDM and WDM runs. Note that the masses of their simulated
systems are larger than the WDM filtering masses used by them. We have shown
in Paper I that the evolution and properties of galaxies formed in WDM halos of masses much higher
than the corresponding \mhm\ are indeed similar to their CDM counterparts.

The plan of the paper is as follows. In Section \ref{method} we present the code, the cosmological background, and
the simulations. In Section \ref{profiles}, we analyze the density and velocity dispersion radial profiles of the 
simulated WDM and CDM dwarfs at $z=0$. In Section \ref{cosmodel} we explore the evolution of the
inner halo density profile of our simulations as a function of several measured quantities in order to 
highlight the mechanism behind the formation or lack thereof of shallow cores. The same phenomena that affect the
inner dark matter distribution can affect the galaxy stellar distribution; we explore this question in 
Section \ref{expansion}. Finally, in Sec. \ref{summary}, we present our summary and final discussion. 

\section{The method}
\label{method}

\begin{table*}[t]
 \begin{center}
  \caption{Physical properties of the runs at $z = 0$}
  \begin{tabular}{@{}ccccccccccc@{}}
  \hline
  Name   & log(\mh)  & log(\ms)   &\rh  & \r2\footnote{Radius where $d\log(\rho)/d\log(r) = -2$.}  &  
log(\ro2)\footnote{Halo density at \r2 .} & 
\con2\footnote{Halo concentration from a local measure: $\con2\equiv \r2/\rh$.}  
&\cnfw\footnote{Halo concentration from globally fitting the profile to the NFW function.}  & \alp2\footnote{Slope of the halo density profile at $0.015-0.02$\rh .} &  \vmax & \re\footnote{Radius that encloses half of the galaxy stellar mass \ms.}\\
& (\msun)  & (\msun) &(kpc) &(kpc)  &(\msun/pc$^{-3}$)  &  &  & & (km $s^{-1}$) & (kpc) \\  
  \hline  
   \multicolumn{10}{c}{ CDM (DM-only)} \\
  \hline
   Dw3    & 10.46 & 8.73 &78.5 (86.1) & 5.1 (4.6)& 15.3 (15.0) & 15.4 (18.9)  &  15.8 (16.9) & -1.46 (-1.70)  & 56.1 (56.8) & 1.0 \\ 
   Dw5    & 10.46 & 8.55  & 77.3 (82.9) & 4.4 (7.9) & 15.5 (15.0) & 17.7 (10.5) & 16.9 (15.3)   & -1.29 (-1.30) & 61.5 (54.6) & 1.1 \\ 
   Dwn1   & 10.63 &  8.90  & 89.8 (97.4) & 7.1 (6.8) & 15.2 (15.2) & 12.7 (14.4) &12.1 (14.7)  &  -1.18 (-1.68)  & 63.7 (61.3) & 2.8 \\ 
   Dwn2   & 10.58 & 8.72  &  84.2 (88.1) & 4.9 (5.5) & 15.4 (15.3) & 17.1 (16.1)  &15.0 (15.9)   & -1.47 (-1.70) & 62.7 (56.7) & 1.3 \\ 
  \hline
   \multicolumn{10}{c}{ \wdma $ $ (DM-only)} \\
  \hline
   Dw3    & 10.43  & 8.74  & 75.2 (83.6) & 8.4 (7.2) & 14.7 (15.0) & 8.9 (11.6) & --- (11.6)   &  -0.66 (-1.10)  & 47.9 (52.0) & 2.0 \\ 
   Dw5    & 10.29  & 7.80  &  68.9 (76.0) & 5.5 (7.6) & 14.9 (14.8) & 12.5 (9.9) & --- (8.9)   &   -0.52 (-1.19)  & 37.6 (47.5) & 1.5 \\ 
   Dwn1   & 10.47  & 8.43  & 78.8 (87.2) & 13.4 (11.1)&14.3 (14.6) & 5.9 (7.8) &  --- (8.2)    &   -0.32 (-1.53) & 43.2 (48.7)& 3.6 \\ 
   Dwn2   & 10.45  & 8.41  & 76.6 (84.5) & 8.5 (11.6) &14.8 (14.5) & 9.0 (7.3) &   --- (8.4)    &   -0.25 (-1.6)& 49.3 (47.3) & 3.9 \\ 
  \hline
 \end{tabular}
 \label{table}
\end{center}
\end{table*}

In Paper I we have presented zoom-in cosmological simulations of four dwarf galaxies
formed in WDM halos that end today with masses close to a filtering
mass \mhm\ (see footnote) equal to $2\times10^{10}$ \msunh, which corresponds to the 
case of a thermal WDM particle mass of 1.2 keV \citep[e.g.,][]{Schneider+2014}. 
The corresponding CDM simulations (same initial conditions) 
were also presented and their results compared to the WDM ones. The
simulations of this latter scenario are the ones 
we analyze here with regard to the halo inner structure and dynamics. 
Below, we briefly summarize the simulation set-up; for more
details, see Paper I. 

\subsection{The Cosmological Background}
\label{cosmology}

Our numerical simulations were run in a cosmology with $\omeb=0.045$, $\ome = 0.3$, $\omel = 0.7$, $h = 0.7$;
the initial power spectrum, $P(k)$, either CDM or WDM, is normalized to $\sige = 0.8$\footnote{The cosmological parameters preferred by the Planck collaboration results \citep{Planck2015} are slightly different from those used by us. As previous authors showed, the parameters that have some effect on galaxy formation and evolution are $\sige$ and $\ome$ \citep[e.g.,][]{Wang+2008,Guo+2013,Henriques+2015}. Our $\sige$ and $\ome$ values are only 3.6\% and 5\% lower than those of the Planck cosmology. These differences would introduce differences in galaxy evolution as a slight delay in structure formation that are much smaller than the uncertainties due to the astrophysical recipes introduced in the simulations. However, our aim here is to explore the qualitative differences produced by baryons on the inner halo and galaxy structure in the CDM and WDM scenarios of galaxy formation. The effects of the cosmological parameters are expected to be much less important than those of the baryon physics and the filtering in the mass power spectrum.}. 
For the CDM case, we use the approximation given by \citet{KH97}, which is good enough for the scales studied in this work. 
For the WDM case, we use the transfer function given by \citet{Viel+2005}
\begin{equation}
T_{WDM} (k) = \left[ 1 + (\alpha k )^{2.0\nu} \right]^{-5.0/\nu},
\label{TF}
\end{equation} 
where $\nu = 1.12$, and $\alpha$ represents an effective free-streaming scale, 
$\lambda_{\rm fs}^{\rm eff}$ \citep[e.g.,][this scale is related to the particle mass, $\mwdm$, $\Omega_{WDM}$, 
and $h$ as given by equation 4 in Paper I]{Schneider+2012}. The transfer function $T_{WDM} (k)$ 
accounts for the deviation from the CDM power spectrum at small scales. Following \citet{Avila-Reese+2001},
we use the half-mode wave number, $k_{\rm hm}$ (that is where $T^2_{WDM}(k)= 0.5$), to 
define our filtering mass, \mhm. At this mass scale is where the abundance and properties
of the halos as well as the properties of the galaxies formed inside them start to deviate significantly from
the CDM case; see Paper I for a discussion and more references therein.

\subsection{The Code}
\label{code}

The simulations were run with the adaptive mesh refinement N-body + Hydrodynamics ART code \citep[][]{KKK97,Kravtsov03}. 
ART includes different astrophysical mechanisms that are relevant for galaxy formation and evolution: atomic and molecular cooling, 
an homogeneous UV heating background source, SF, metal advection, and thermal feedback. 
The cooling and heating rates are tabulated for a temperature range of $10^2 < T < 10^9\ \Ke$ and a grid of densities, 
metallicities (from $Z = -3.0$ to $Z = 1.0$, in solar units), and redshifts using the CLOUDY code \citep[version 96b4]{Ferland98}.

For a discussion on the subgrid physics used in our simulations, we refer the reader to 
\citet{Colin+2010} and \citet[][see also \citealp{Avila-Reese+2011}]{Gonzalez-Samaniego+2014}.
In Paper I we have used the same subgrid parameters for SF and feedback processes as in \citet{Gonzalez-Samaniego+2014}. 
In particular, SF takes place in those cells for which $T < \Tsf$ and $n_g > \nsf$, where $T$ and $n_g$ are the 
temperature and number density of the gas, respectively, and $\Tsf$ and $\nsf$ are the corresponding threshold values.
A stellar particle of mass $m_* = \esf\ m_g$ is placed in a grid cell every time the above conditions are simultaneously 
satisfied, where $m_g$ is the gas mass in the cell and \esf\ is a parameter that measures the local efficiency
by which gas is converted into stars. We set $\Tsf = 9000$ \Ke, $\nsf = 6$ \cmtres\ and $\esf = 0.5$
(\citealp[see][]{Avila-Reese+2011} and \citealp[][for a justification of these values]{Gonzalez-Samaniego+2014}).

Our feedback implementation works in such a way that each star more massive than 8 \msun\ injects instantaneously
into the stellar particle location $E_{{\rm SN+wind}} = 2 \times 10^{51}\ \erg$ of {\it thermal} energy.{\footnote{ The sudden injection of energy is able to raise the temperature of the gas in the cell to several $10^7$ K, high enough to make the cooling time larger than the crossing time and thus avoiding most of the overcooling.}  
Half of this energy is assumed to come from the type-II SN and half from the shocked stellar winds.
Each one of these massive stars dumps $1.3 \msun$ of metals into the interstellar medium. ART accounts 
for SNIa feedback assuming a rate that peaks at 1 Gyr. It also injects $ 2 \times 10^{51}\ \erg$ for each SN event.
In our previous \lcdm\ simulations of low-mass galaxies, a cooling delay was imposed for a while on the cells where the
young stellar particle is to reduce overcooling, arising in part due to resolution issues. For a fair comparison between 
the WDM runs and their CDM counterparts, we have kept this delay; though, as discussed in Paper I, the
simulation results at our resolution are hardly affected if cooling remains. Our stellar feedback recipe is justified in the end as an effective way to increase its strength. The scheme resembles the one implemented in other simulations,
e.g., the EAGLE simulation \citep{Schaye+2015}.

\subsection{The simulations}
\label{simulations}

As explained in Paper I, we focus our analysis on galaxies formed in WDM halos of masses close to the filtering scale \mhm. In particular, 
we analyze here four zoom-in WDM simulations that end with virial masses around the filtering mass, $\mhm=2\times 10^{10}$ \msunh.
For each one of these simulations, we have run their CDM counterparts
with the same initial conditions and subgrid physics. The simulations called Dw3 and Dw4 in the \lcdm\ scenario 
were studied in \citet{Gonzalez-Samaniego+2014}, while Dwn1 and Dwn2 were first
identified in the \lwdm\ cosmology and then simulated in the \lcdm\ cosmology. The halo and stellar mass assembly histories of these four runs were presented in Paper I (Figure 5 therein). The halo histories are different: from Dw3 to Dwn2 they go from earlier assembly to one more extended in time. These differences are more noticeable in the \lcdm\ runs. In the \lwdm\ case, the assembly of  the bulk of the mass is shifted to lower redshifts in all runs. The halos in both cosmologies do not suffer major mergers since $z\sim 1$, excepting Dwn2. In general, the \lwdm\ halos suffer fewer mergers than the \lcdm\ ones.

The galaxy/halo systems re-simulated at high resolution were selected as isolated halos in an N-body simulation of a cosmological 
box of $L_{box}=10\ \mpch$ side and a root grid of $128^3$ cells. These halos, both in the \lcdm\ and \lwdm\ cosmologies, do not have halos of similar or larger masses at distances smaller than $\sim 1$ \mpch\ and they reside in local low-density regions.
The halos are characterized by its virial radius, which is defined 
as the radius that encloses a mean density equal to $\Delta_{vir}$ times the mean density of the universe, where $\Delta_{vir}$ is 
obtained from the spherical top-hat collapse model. The halos at this radius end up with more than half a million DM particles 
in the high-resolution zone.

In order to avoid the effects of a spurious structure formation in the \lwdm\ simulations, we have used a less aggressive refinement in 
these runs with respect to the \lcdm\ ones. In the Appendix of Paper I we have checked that this does not introduce a bias in
the \lwdm\ simulations with respect to the  \lcdm\ ones. In any case, the WDM halos end up with masses at $z=0$ 
(or even at $z\sim 2$) much larger than the scale where artificial structures are expected to form according to 
\citet{Wang+2007}. The proper size of the finest grid cells in the simulations changes with the scale factor, being
at the onset of SF, $z \sim\ 3$, around 55 pc and at $z=0$, $\sim 110$ pc.

 In Table 1 we summarize the $z = 0$ properties of our four simulated halos/galaxies in the WDM and CDM cosmologies, both for
the N-body dark matter only simulations (within parentheses, when applicable) and for the hydro simulations that include baryons.
The stellar mass of the galaxy \ms\ is defined as the mass 
in stellar particles within 0.1\rh, and the stellar effective or half-mass radius \re\ is the radius where half of 
\ms\ is contained.

\section{Density and velocity dispersion radial profiles}
\label{profiles}

Do the mass density profiles of the \lcdm\ and \lwdm\ halos differ significantly? 
Let us first analyze the {\it dark matter only} (N-body) simulations. The
black short-dashed and solid lines in Fig. \ref{fig1} show the density profiles of the four halos
in the \lcdm\ and \lwdm\ cosmologies, respectively.   
Actually, differences are small. In the center both have roughly
the same slope and overall the CDM halos look more concentrated than the WDM ones. 
We have fitted the measured density profiles to the conventional Navarro-Frenk-White (NFW) profile \citep{Navarro+1997}.
The obtained close-to-one reduced $\chi^2$'s (typically $1.1-1.3$ and not larger than 1.7) show that the 
NFW profile describes well both the CDM and WDM simulated halos. 
However, the NFW concentration parameters, \cnfw, are higher for the CDM halos than for the WDM ones.
From the Dw3 to the Dwn2 run, the difference increases systematically from factors of $\approx 1.5-1.9$ (see Table 1, quantities in parentheses). Therefore, we confirm the results of previous works
(see the references in the Introduction) concerning the differences 
in concentration between the CDM and WDM halos.
Note that our simulations here are for WDM halos of masses around the corresponding
filtering scale, where strong evolutionary differences with the CDM halos are expected. 

 As a sanity check, we have also calculated the halo concentration \con2\ defined as \rh/\r2, where
\r2\ is the radius where the logarithmic density slope is $-2$. Thus, while \cnfw\ is obtained for a fit to the overall
density profile, \con2\ is determined from a local measure. For the analytical NFW profile, both concentrations
are equal by definition.  For the simulation results, both concentrations do not differ by more than 15\%,
except for the Dw5 CDM run, which presents a little bump in its density profile around the NFW
scale radius, causing the $-2$ numerical slope to be moved to a larger radius.  

\begin{figure}
\plotdelgado{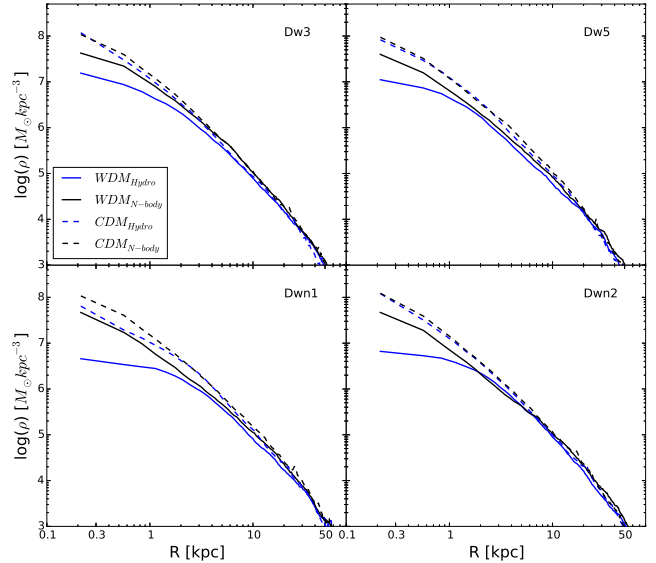}
\caption{Dark matter density profiles at $z=0$ for the WDM (solid lines) and CDM (dashed lines) runs.  The blue color is for the
hydro simulations while the black color is for the dark matter only simulations. }
\label{fig1} 
\end{figure}

To quantify the inner mass distribution of the halos, we measure 
the slope of the density profiles in the bin [0.015,0.02]\rh\ by
assuming a power law $r^{\alp2}$ fit for this interval. 
For all of our runs, $0.02$\rh\  is in between 
1.1 and 1.8 kpc, well above the nominal resolution of our simulations. For the CDM runs, $\alp2$ is in
between -1.3 and -1.7 in
agreement with the slope of $-1.4$ obtained at $0.02$\rh\ for an NFW profile corresponding to a $3\times 10^{10}$ \msun\
halo with $\cnfw\approx 15$. For the WDM runs,  $\alp2$ is in between -1.1 and -1.6, also in agreement with an NFW profile
of the same mass but with $\cnfw\approx 9$. 
The densities at \r2, $\rho_{-2}$, are lower in the WDM halos than in the CDM ones.  
In summary, the density profiles of our WDM dwarf halos in the DM-only simulations 
{\it are reasonably well described by the NFW function, but with respect to their CDM counterparts
they are less concentrated by factors $1.5-2$ and slightly less cuspy.} Thus, the inner gravitational potential
in the WDM halos should be less deep than in their CDM counterparts. 

Let us now analyze the hydro simulations which include baryons. The blue
short-dashed and solid lines in Fig. \ref{fig1} show the halo density profiles 
of the four runs in the \lcdm\ and \lwdm\ cosmologies, respectively.
Notice that in the hydro simulations not all the mass is in dark matter; initially 15\% of the total mass is in baryons
(however, a fraction of them, up to $\sim 80\%$ can be lost during the evolution
due to galactic winds). Therefore, the dark halo density
profiles in the hydro simulations should lie slightly below those of
the N-body simulations. We see that while 
the effects of baryons on the density profiles of the CDM halos are modest, in the case of the WDM halos, they produce 
a significant flattening of the inner density profile. 

The CDM halos in the hydro simulations are still well fit by the NFW profile (the reduced $\chi^2$'s
are between 1.1 and 1.3). From Table 1 we see that the \cnfw\ values differ from those of the dark matter only simulations 
by less than $\approx 20\%$, and the inner slopes \alp2\ are roughly the same after the inclusion 
of baryons.  Therefore, for our subgrid scheme and parameter values we use in our simulations, {\it the effects of baryons 
are not able to produce shallow cores systematically in the CDM halos of $\mh=2-4\times 10^{10}$ \msun} presented here
and in most of the halos presented in \citet[][]{Gonzalez-Samaniego+2014}.\footnote{In \citet{Gonzalez-Samaniego+2014}, seven CDM simulated dwarfs
with different halo mass assembly histories were presented (two are the Dw3 and Dw5 analyzed here). While in five of them shallow 
cores do not form, in the other two, which have very late and extended assembly histories, the inner 
density profile tends to flatten; see the discussion in Section \ref{summary} below.} 

In the case of the WDM halos, shallow halo cores are formed systematically.  
From Table 1, the measured slopes \alp2\ are clearly shallower in the WDM runs than in the CDM ones, with 
values between $-0.25$ and $-0.66$, much flatter than those from the NFW profile. Therefore, the NFW profile 
does not provide a good fit to the dark matter halos in the \lwdm\ hydro simulations. 
Regarding the overall mass distribution, we use the \con2\ concentration parameter to compare the WDM and 
CDM halos in the hydro simulations. As seen in Table 1, the former are less concentrated than the latter
by factors $1.4-2.2$, and the $\rho_{-2}$ densities are $0.6-0.9$ dex lower. 

By comparing the \con2\ parameter of the WDM halos in the N-body and hydro simulations (see Table 1),
we see that the effects of baryons make the halos in two cases less concentrated than without them,
and in the other two cases make the halos more concentrated, the differences being less than $\sim 30\%$
in all cases. Summarizing, {\it the baryons in our WDM halos of masses around \mhm\ promote
a significant flattening of the inner density profiles}, without affecting in a systematic way the overall 
mass distribution of the halos. 

\begin{figure}
\plotdelgado{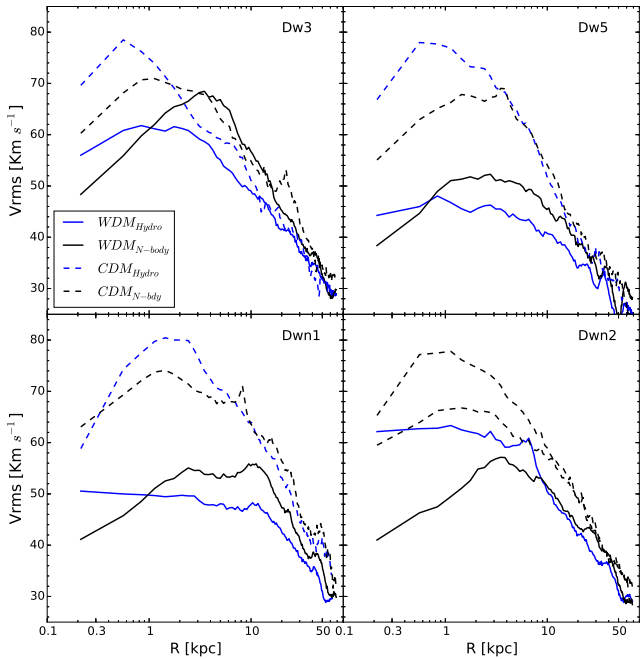}
\caption{Dark matter 3D velocity dispersion profiles at $z=0$ for the WDM (solid lines) and CDM (dashed lines) runs.  The blue color 
is for the hydro simulations while the black color is for the dark matter only simulations.}.
\label{fig2}
\end{figure}

In Fig. \ref{fig2}, the $z=0$ 3D velocity dispersion ($\sigma_{\rm 3D}$) profiles corresponding 
to the same four WDM and CDM halos as in Fig. \ref{fig1} are plotted. For the dark matter only 
simulations, the velocity dispersions of the CDM halos (black dashed lines) are higher than those of the 
WDM halos (black solid lines), especially toward the central regions. This is consistent with the fact that
the WDM halos are significantly less concentrated than the CDM ones, which, according to the Jeans
equation, implies lower velocity dispersions to attain equilibrium.  The $\sigma_{\rm 3D}$ profiles, both for the CDM and WDM 
cases decrease toward the center from a maximum attained at radii smaller than the corresponding
NFW scale radius. This is the behavior expected for halos with an NFW density profile in
virial equilibrium and with orbits close to the isotropic case \citep{Lokas+2001}.

For the hydro simulations, the CDM halos show $\sigma_{\rm 3D}$ profiles (blue dashed lines)
with similar shapes as in the case of the dark matter only simulations, but with the peaks shifted to 
smaller radii and with slightly higher values. This suggests slightly more anisotropic 
orbits in the center, likely due to the effects of baryons. In the case of the WDM halos, the velocity dispersion 
profiles tend to flatten in the inner halo regions, evidencing a strong dynamical rearrangement of the particles.

\begin{figure}[h!]
\plotdelgado{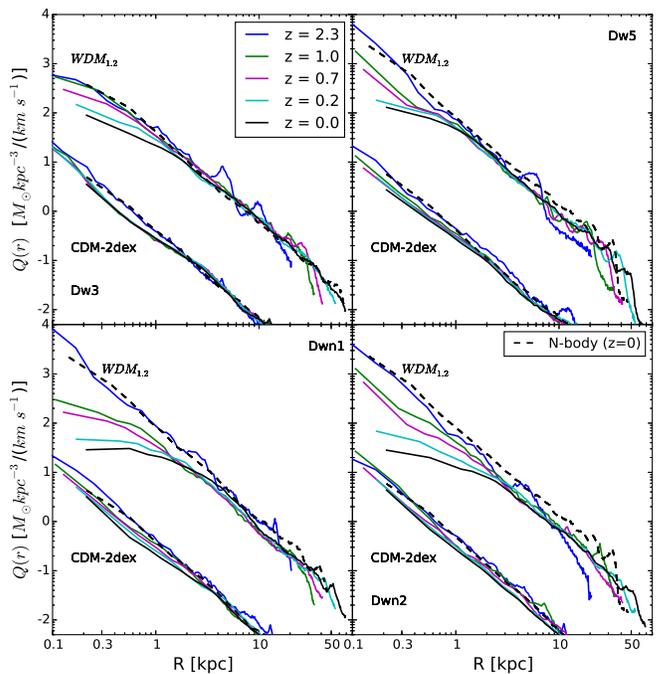}
\caption{Coarse-grained pseudo phase-density profiles at different redshifts for the WDM and CDM runs (solid curves); the latter are shifted down by 2 dex. 
The different colors correspond to the different $z'$s indicated with labels inside the upper left panel. The black dashed lines correspond to the $z=0$
profiles in the corresponding dark matter only simulations.}
\label{psdensity}
\end{figure}

\section{Evolutionary trends and shallow core formation}
\label{cosmodel}

In the previous section we have seen that shallow and nearly ``isothermal'' halo cores
form at $z=0$ in our WDM hydro simulations.
How is the dynamical evolution of these systems? Do they have established shallow halo
cores since early epochs or these cores form gradually? 

An important property of collisionless systems is that the phase space density of the fluid
stays constant with time. Since it is not possible to measure the phase space density in 
the outcomes of numerical simulations, one uses the coarse-grained phase space density instead; 
that is, we define the mass density $F({\bf x, v}; t)$,
in a discrete six-dimensional phase space volume, 
$\Delta^3${\bf x}$\Delta^3${\bf v}, centered on the point ({\bf x,v}) at time $t$.  A proxy for this quantity 
is the coarse-grained pseudo phase-space density, defined as $Q\equiv \rho/<\sigma_{\rm 3D}^2>^{3/2}$
\citep{Hogan+2000}. The shape of the radially averaged  $Q(r)$ profile measured in the 
simulated halos is similar to that of the radially averaged coarse-grained phase space density, but its 
amplitude is shifted to larger values,  $Q(r)>F(r)$ \citep{Shao+2013}. For CDM halos, $Q(r)$ 
is well fitted by a single power law of slope $\approx -1.9$ \citep[e.g.,][]{Taylor+2001}. 
For WDM halos, a flattening in the inner $Q(r)$ as well as the $F(r)$ profile and the existence of an upper
limit are expected, given that the WDM particles have non-zero thermal velocities; this is a 
consequence of the Lioviulle's theorem \citep[e.g.,][]{Dalcanton+2001}. Therefore, given that the velocity dispersion 
does not grow in the inner region of a halo, the real-space density profile must become constant with 
a core size depending on the specific WDM model \citep{Tremaine+1979}. For WDM thermal particles 
of masses $\sim 1$ keV, the corresponding thermal relict velocities are so low that the implied 
cored regions would be too small to explain the shallow cores suggested by some 
observational studies, as was first shown by \citet{Avila-Reese+2001} and \citet{Colin+2008}.

In Fig. \ref{psdensity}, the $Q(r)$ profiles for our four WDM and CDM
halos in the hydro simulations are shown at different redshifts (color
solid lines). The black dashed lines are for the dark matter only simulations at $z=0$. 
While for the CDM halos the pseudo phase-space density profiles remain
almost the same and are well described by a power law with slope $-1.9$, for the 
WDM halos there is a clear trend of flattening $Q(r)$ in the center with time. 
For the latter, it is interesting to see that $Q(r)$ profiles at the epochs of halo formation
are very close to the $Q(r)$ profiles of the $z=0$ dark matter only simulations and
well described also by a power law with slope $-1.9$ (dashed black lines). 

The inner flattening of the pseudo phase space density profiles in the WDM hydro
simulations suggests that the shallow halo cores are produced by {\it the dynamical heating 
of the dark matter particle orbits in the central regions with the consequent expansion 
of these regions.} Why are the processes that drive the expansion of the core 
more efficient in the \lwdm\ hydro simulations 
than in the \lcdm\ ones? In the following section, we explore this question in more detail.

\subsection{Flattening of the inner halo density profile}

A way to quantify the evolution of the inner part of the halos is by measuring 
the change with time of the inner slope of the density profile at 
some scaled radius, 
for example, in the bin $0.015-0.02\rh$, \alp2.  Figure \ref{fig5} shows \alp2\ 
measured at many epochs for our four WDM and CDM halos (upper and
lower panels, respectively) in the hydro simulations.
The trends are clear: while in the WDM halos a systematic flattening
of the inner density profile is seen, in the CDM halos the cores become slightly more cuspy with time. 
Note that at early times
the inner bins, where \alp2\ is measured, are resolved with less particles,
so at these epochs the reported values are uncertain. 

It is important to note that in the \lwdm\ scenario the overall formation process 
of structures of scales close to \mhm\ is delayed. According to Fig. \ref{fig5}, 
$\alp2\approx -1.1\div -1.4$ at the epochs when the halos start to significantly 
assemble ($z\simless 2$ or ages $\simgreat 3.3$ Gyr). At these epochs, the massive 
transformation of gas into stars is just about to start; the \ms-to-\mh\ ratios are 
very low, $<10^{-3}$. At later times, all the halos show a systematic 
increasing of the \alp2\ values (flattening of the inner density profile), in each case 
with a different rate, and during the last 3-5 Gyr, \alp2\ oscillates around nearly constant values, 
$\approx-0.3$ for Dwn1 and Dwn2, and $\approx-0.45$ for Dw5. Only in the case of 
Dw3, \alp2\ decreases a little after a maximum and since $z\sim 1$ it remains nearly constant at a value of 
$\approx -0.7$; this is the WDM run that ends with the least shallow core among the four runs. 

The halos and galaxies in the \lcdm\ scenario start to assemble at very early epochs. 
At redshifts $\sim$4--2 the SF is very active in these galaxies when the halo assembly is in its fast regime.
According to Fig. \ref{fig5}, at these epochs the CDM halos have values of \alp2\ as shallow as 
$\approx -0.7\div -1.1$. At lower redshifts, when the SFR has typically decreased, 
the inner slopes \alp2\ become quickly steeper (likely because the inner halo slightly 
contracts due to the concentration of baryons) and remain roughly constant 
for the next $\sim 7$ Gyr, with values as steep as $-1.2\div -1.5$.

\begin{figure}
\plotdelgado{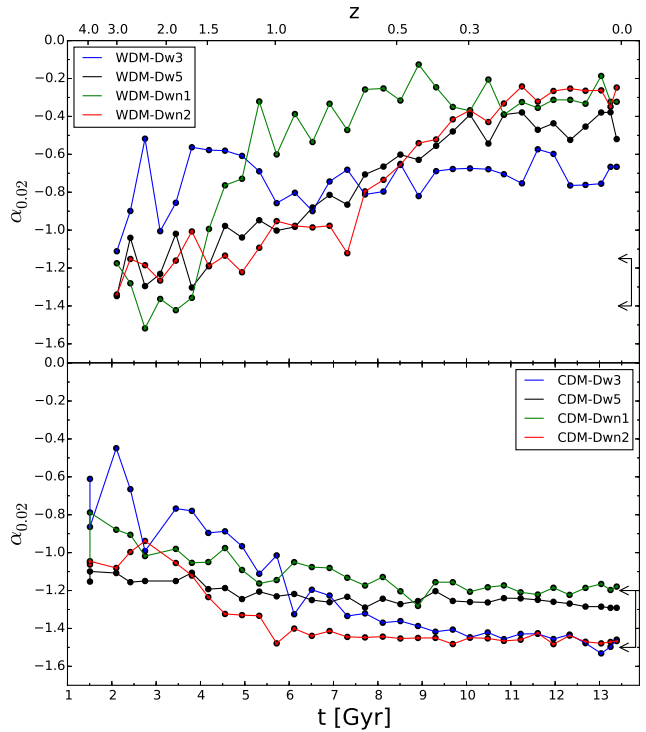}
\caption{ Evolution of the inner slope (at $R\approx 0.02\rh $) of the density profiles, \alp2, for the four simulated dwarfs in both
the WDM (upper panel) and CDM (bottom panel) cosmologies. The arrows indicate the values of \alp2\ corresponding
to NFW profiles with concentrations in the range measured at $z=0$ in our WDM (upper panel) and CDM (lower panel) 
dark matter only simulations (see Table 1 for the values of these concentrations)}.
\label{fig5}
\end{figure}

A key quantity of galaxy evolution is  the time-integrated SF efficiency of halos,
measured through the ratio of \ms\ to \mh. 
For our four \lwdm\ dwarfs, we see that when at a given epoch the inner slope \alp2\ is 
shallower than $\approx -0.5$,  $\log$(\ms/\mh) is then higher than $-2.6$ and lower than $-2.0$.
However, for values of $\log$(\ms/\mh) in the same range, values of \alp2\ as steep 
as $-1.3$ also can be found (at redshifts $z\simgreat 1$).  
For $\log$(\ms/\mh)$<-3$, \alp2\ at any epoch is steeper than $-0.9$ for the four halos. 
Therefore, shallow cores are not formed for too low integral SF efficiencies. 
Thus, it seems that {\it a necessary (but not sufficient) condition for inner halo density profile flattening 
at any epoch is that the \ms-to-\mh\ ratio of the system had values in the range $0.25-1$\%,}
in agreement with the results by \citet[][]{diCintio+2014} \citep[see also ][]{Tollet+2015}. 
For our subgrid physics, the \lcdm\
simulations of the systems that end today with masses $\approx 2-4\times 10^{10}$ \msun,
have $\log$(\ms/\mh)$>-2$ at $z=0$, so their halos do not present shallow cores. On the
other hand, 
the corresponding \lwdm\ simulations have $\log$(\ms/\mh) values between $-2.4$ and $-2.0$
at $z=0$, except for Dw3, which has a higher value. The inner slope of this run at $z=0$
is $\alp2\approx -0.7$, which is the steepest of all of the runs (see Table 1).

It is not our aim here to perform exhaustive comparisons of the results with inferences of inner halo mass distributions
    from observations, which are still ambiguous. Several authors attempted to constrain the inner density slope of dwarf and LSB galaxies from the observed kinematics. Depending on the resolution of the observations and under several assumptions, some authors report inner density slopes for most of the analyzed galaxies shallower than $\sim -0.7$ at radii between $\sim 0.1$ and 1 kpc. At larger radii, the slopes are steeper \citep[e.g.,][]{deBlok+2001,deBlok+2002,Oh+2015}. The inner slopes measured in our WDM runs at $z=0$ between the conservative resolution radius of $\sim 0.7-0.8$ kpc and the radius where the profile starts to significantly flatten are $-0.65<\alpha<-0.2$, in good agreement with those observational inferences that have resolution radii $R_{\rm in}\sim 0.7-1$ kpc. 

\begin{figure}
\plotdelgado{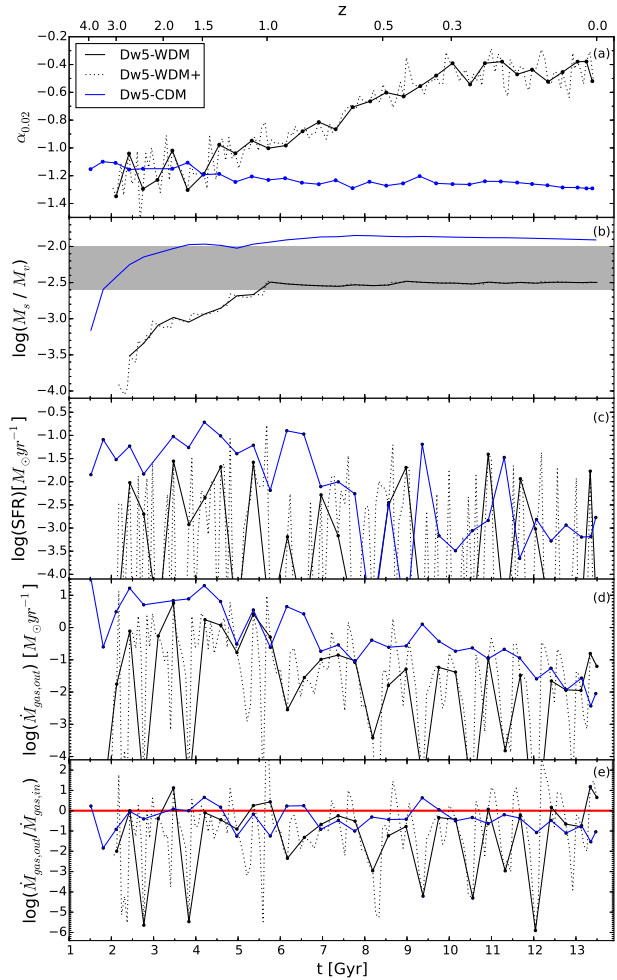}
\caption{ From top to bottom, (a) evolution of the inner slope \alp2, (b) the \ms-to\mh\ ratio, (c) the SFR, (d) the galaxy gas outflow rate, and (e) the
galaxy gas outflow-to-inflow ratio, for the Dw5 runs. Solid black and blue curves are for the WDM and CDM cosmologies, respectively. 
The dotted line in each panel is for the same WDM run but including more time snapshots.}
\label{mixevol}
\end{figure}

\subsection{Density Profile Flattening vs. Galaxy Evolution}

So far, we have seen that the formation of shallow cores {\it at any time} in our hydro WDM 
simulations happens only when the \ms-to-\mh\ ratio has values between 0.25\%\ and 1\%.
Are there other galaxy properties that could be related to shallow core formation?

To explore the above stated question, we make a case study of the evolution of the Dw5 system in both 
the WDM and CDM cosmologies; the other runs have similar behaviors in general.
In Fig. \ref{mixevol} we present the evolution of the \alp2\ inner slope of this run for both 
the CDM and WDM cosmologies (panel (a)) along with the evolution of the respective
\ms-to-\mh\ ratios, SFRs, gas outflow rates from a sphere of radius 0.1 \rh,
and the gas outflow-to-inflow ratios at the same radius (panels (b)--(e), respectively).
The black and blue solid lines are for the WDM and CDM cases, respectively. 
For the WDM run, we have also computed the evolution of all of the abovementioned quantities 
from a catalog with more snapshots,\footnote{From $a=0.2$, timesteps are saved every
$\Delta a=0.005$, where $a$ is the expansion scale factor of the universe.
In total, this simulation has 168 snapshots.} Dw5-WDM+, and plot their results with dotted lines.

As seen in panels (a) and (b), the inner slope \alp2\ of the WDM run starts
to increase significantly since $z\sim 1$, the epoch when $\log$(\ms/\mh) increases up to the value
of  $-2.6$. In the case of the CDM run, \alp2\ is shallower than $-1.15$ at $z>1.7$,
when $\log$(\ms/\mh) is lower than $-2$. The SFR histories plotted in panel (c) 
are computed as the ratio between the amount of gas transformed into stars in a time interval 
$\Delta t$ and this time bin, at the given epoch. For $\Delta t$ we chose 100 Myr,
which is a little longer than the life of a star of eight solar masses. The SFR is strongly episodic in 
the WDM case; this can be appreciated particularly when the time resolution between the snapshots 
is increased (dotted line).  The galaxy had the most intense bursts of SF at times between $\sim 2$ 
and 5.5 Gyr; at these epochs, \alp2\ is steeper than $-1.0$, but just around 
the age of 5 Gyr the strong flattening of the core starts. Note that the SFR decreases significantly 
namely when this happens. Then, since $\sim 7$ Gyr new strong bursts of SF happen, though
they are mostly of lower amplitude than in the initial 5.5 Gyrs. In the case of the CDM run, the SFR
is higher most of the time than the WDM one but it is significantly less bursty (comparing 
results with the same number of snapshots).  The early mild flattening of the core in the CDM
dwarf happens in the epochs of very active SF.

In summary, for both the WDM and CDM runs, we do not see a correlation between the average 
amplitude of the SFR and the \alp2\ inner slope. However, {\it the strong intermittence of the SF at
all epochs in the WDM run could be related to the halo core flattening. All the WDM runs show
this behavior.} Several authors have reported that shallow cores form in their \lcdm\ cosmological
simulations when the SFR strongly fluctuates and it is sustained \citep[e.g.,][]{Governato+2012, 
Teyssier+2013, Madau+2014,Onorbe+2015, Chan+2015}.

In fact, rather than the SF, the SN-driven outflows are the ones that could be directly associated
with the halo core flattening, according to the impulsive blow-out scheme (\citealp{Navarro+1996};
\citealp{Read+2005}). The outflow rates measured at 0.1\rh\ (roughly the galaxy extent) for the Dw5 
dwarf are plotted in panel (d). For the WDM case, the outflow rates are very episodic and strong
at early times ($2\simless$ t/Gyr $\simless 6$), when the inner slope \alp2\ is mildly 
changing. The outflow rate histories of the WDM and CDM runs are qualitatively similar, though the
former are more episodic and have lower amplitudes than the latter 
(comparing results with the same number of snapshots). 
At later epochs, the outflow rates decrease in both runs. 
As expected, there is a correlation between the SF and mass outflow rates, though in the CDM case, a less clear correlation is seen at late times;
in particular sometimes the strong SFR drops are not followed by strong drops in the mass outflow rates.  Yet, if we repeat the analysis
for a larger radius, say 0.2\rh, we recover a good correlation; that is, drops in SFR correlate with drops in the outflow rate. It 
appears that in the CDM run, at late times,  the gas can overcome the 0.1\rh\ radius sphere but it does not reach already the radius 0.2\rh. 

Again, as with the SFR history, we also do not observe a correlation between inner density profile 
flattening in the WDM  run and the overall outflow rate history.  Actually, during the evolution of the 
halo/galaxy system, along with the outflows, an inflow of gas is also present. A significant
fraction of the accreted gas transforms into a body of stars in the center of the
system, thus deepening the local potential well. In panel (e) we plot the evolution 
of the outlow-to-inflow ratio at 0.1\rh\ for the same WDM and CDM Dw5 runs as in 
the other panels. As can be seen, most of the time this ratio is lower than 1, that is, gas accretion
rather than gas ejection dominates. The former is actually happening smoothly and 
decreases with time. The latter, as already shown, is episodic (specially in the WDM run),
so that the shape of the ratio evolution is completely dominated by that
of the outflow rate evolution. 

It has been proposed that the main reason for halo profile flattening is
the intense and rapid variations with time of the gas content in the very central regions 
\citep{Mashchenko+2006,Pontzen+2012}. These variations can induce quick fluctuations 
(on time scales shorter than the dynamical time) in the central gravitational potential 
and are associated with rapid movements of the center of mass of the gas with respect to that 
of the dark matter; both effects seem to be an efficient mechanism for transferring kinetic energy from
the gas to the dark matter, causing the inner halo (and stellar) regions to expand.
 
In Fig. \ref{fluctuations} we plot the change with time of different masses contained inside 500 pc proper around the 
halo center of the same Dw5 run shown in Fig. \ref{mixevol}: dark matter ($M_{\rm dm}^{500}$; black line), stars 
($M_{\rm s}^{500}$blue line), gas ($M_{\rm g}^{500}$; gray line), and the sum of all the components ($M_{\rm tot}^{500}$;
red line). Upper and lower panels are for the WDM and CDM cases, respectively.
For the WDM simulation, we used the catalog with more snapshots, which was not available for the respective CDM run.
To compare results from both cosmologies, in the lower panel we also plot the WDM results but from the catalog with 
the same number of snapshots as the CDM one (dotted lines).

As expected, the change with time of $M_{\rm dm}^{500}$ follows the same trend as \alp2\ for both the WDM and CDM runs. 
For the former, in the time period where \alp2\ gets significantly shallower ($5 \simless$ t/Gyr $\simless 10$ Gyr),
$M_{\rm dm}^{500}$ decreases by a factor of $4-5$, with small fluctuations. Until $t\approx 5.5$ Gyr, the amount of mass 
in stars within 500 pc increases to values similar to those in dark matter, and then both decrease, though in the last $\sim 4$ Gyr,
$M_{\rm s}^{500}$ decreases more than $M_{\rm dm}^{500}$ and with variations on average less than or equal to a factor of three. 

The gas mass inside the inner 500 pc of the halo is not larger on average than $\sim 1/3$ of the total mass during the epochs when the 
flattening of the inner halo profile happens, but it strongly fluctuates on very short timescales (less than the 
time interval between consecutive snapshots, which is of $50-80$ Myr). The fluctuations have amplitudes of 5--15 on 
average but can be as large as 100--200. Therefore, while subdominant in mass, the gas content within 500 pc fluctuates 
so much and so quickly that this surely produces non-negligible fluctuations in the central gravitational potential; 
besides, most of the strong variations in gas mass within the inner 500 pc of the dark matter halo are due to 
bulk motions of the gas.\footnote{We have seen that the centers of mass of the gas and dark matter 
components move back and forth constantly most of time and in all the runs. These off-centers are typically 
of scales close to the size resolution of the innermost halo bin but sometimes they are larger.} 
{\it These are namely the conditions proposed for dynamical heating of the dark 
matter particle orbits}. During the last $\sim 4$ Gyr, when the slope \alp2\ fluctuates significantly but around a 
constant value of $-0.45$ on average, the amplitude of the gas mass variations is remarkably lower on average 
than at earlier epochs. Note that at late epochs, the gas mass content becomes of the order of or larger 
than that of the dark matter and stellar mass contents.

For the CDM run (bottom panel), $M_{\rm g}^{500}$ does not vary significantly with time since
$t\approx 4.5$ Gyr, at least not as in the WDM run (dotted lines in this panel). The most intense variations in 
$M_{\rm g}^{500}$ happen at early epochs when the slope \alp2\ is slightly shallower than later. 
In general, the central region of the Dw5 CDM halo shows a passive evolution, with roughly constant 
amounts of dark matter, stars, and gas. Therefore, no shallow core formation is expected for this galaxy. 

\begin{figure}
\plotdelgado{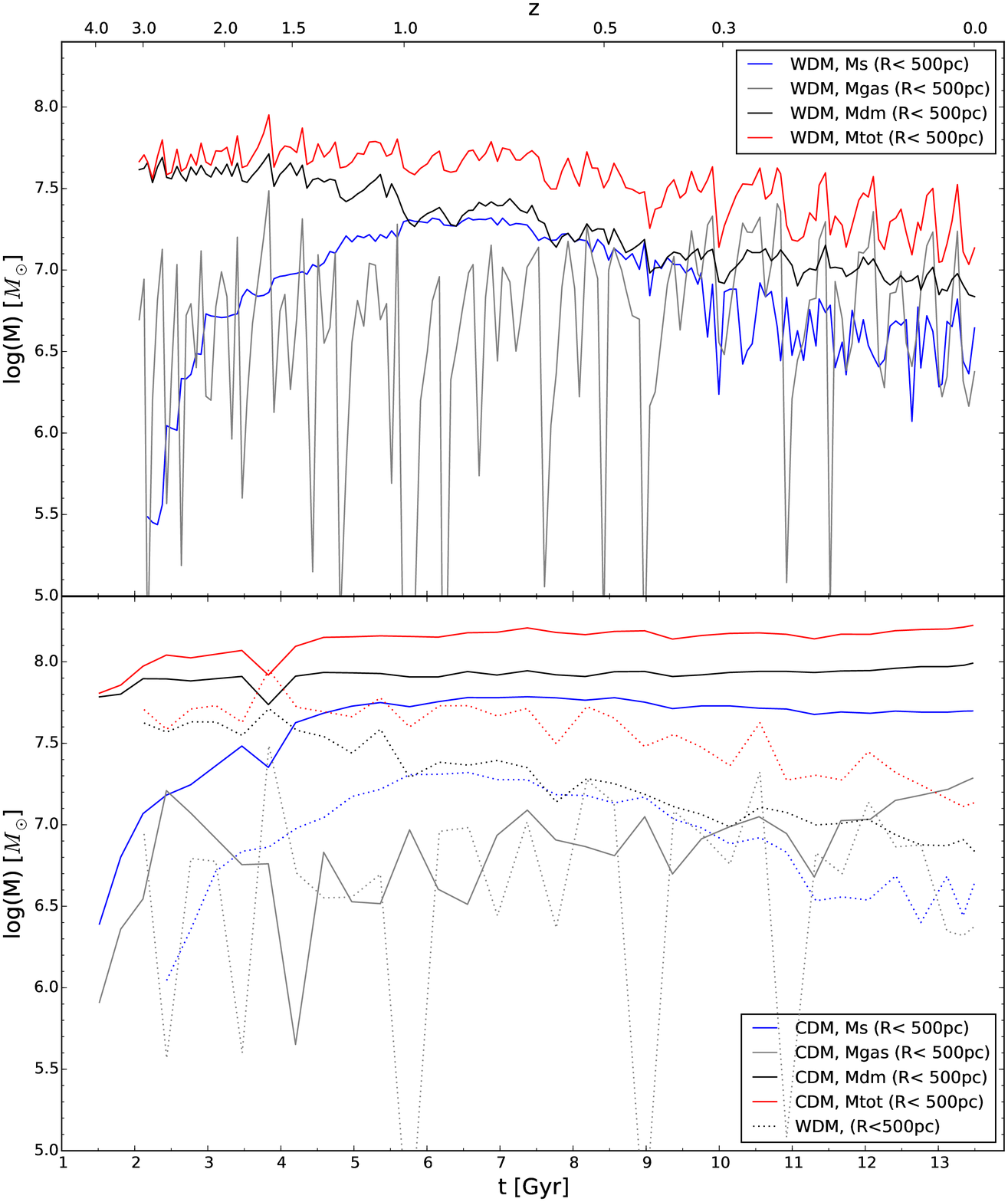}
\caption{Evolution of enclosed mass within the halo central 500 pc for stars (blue), gas (gray), dark matter (black), and total mass (red) in the Dw5 run
for the WDM (upper panel) and CDM (lower panel) cosmologies. For a fair comparison between the WDM and CDM runs, in the bottom panel the 
results for the WDM case with the same amount of time snapshots (dotted curves) as we used for the CDM run are plotted with dotted lines.}
\label{fluctuations}
\end{figure}

\begin{figure}
\plotdelgados{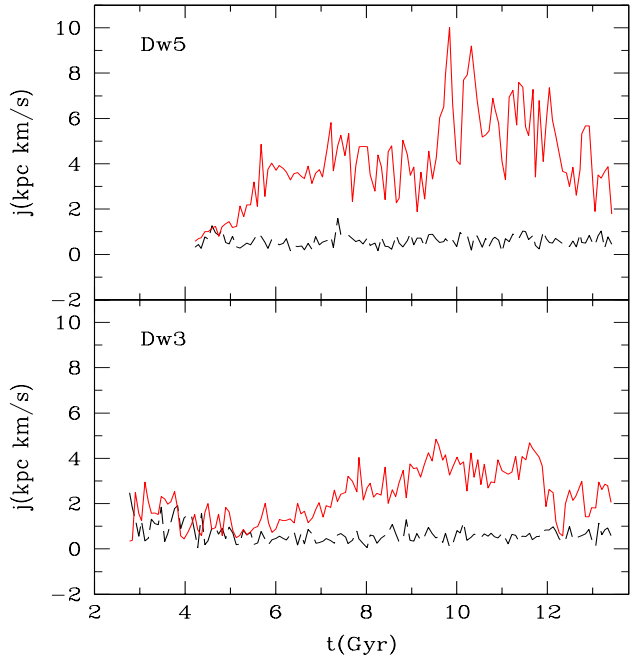}
\caption{Evolution of the specific angular momentum for stars (red), and dark matter (black) particles within $R < 1.0$ kpc,  in the Dw5  and Dw3 runs, top and bottom panels respectively, for the WDM cosmology.}
\label{jdm}
\end{figure}

\subsection{Angular Momentum Transference to Dark Matter}
Finally, we explore the possibility that the flattening of the dark matter profile in the WDM hydro simulations
may be due to the transfer of angular momentum from baryons to dark matter. This can happen as a result of the
dynamical friction between stellar clumps or a possible stellar bar and the inner dark matter background 
\citep[e.g.,][]{Romano-Diaz+2008, delPopolo+2009, Valenzuela+2003}. In Fig. \ref{jdm} we show the evolution of the specific angular 
momentum of the dark matter and stellar components ($j_{\rm DM}$ and $j_s$, respectively), inside a sphere 
of radius 1 kpc proper, centered on the galaxy, for the runs Dw3 and Dw5.  We find that $j_{\rm DM}$ for both dwarfs 
remains nearly constant in time with about the same value, around $0.5\pm 0.2$ kpc km s$^{-1}$. Instead,
$j_s$ presents an episodic evolution in both runs, increasing significantly with respect to the very early values,
which are similar to those of the dark matter (by factors $\sim 4-5$ and $\sim 6-10$ for Dw3 and Dw5, respectively).
In conclusion, {\it the gain/loss of specific angular momentum of the stellar particles does not correlate 
with the specific angular momentum of the dark matter particles in the central halo regions.}

\section{Galaxy expansion and stellar migration}
\label{expansion}
 
As seen in Fig. \ref{fluctuations},  for the WDM Dw5 galaxy, not only the dark matter mass in the innermost 500 pc proper of the halo region
decreases with time but also the stellar mass (since $t\sim 6.5$ Gyr) decreases with time; the latter 
decreases even faster during the last  3.5 Gyr. 
This suggests that the dynamical processes that expand the inner dark matter halo regions, producing a shallow core, 
also affect the collisionless stellar component as discussed in \citet{Read+2005} and \citet{Maxwell+2012}, and reported 
in cosmological simulations of galaxies in \citet{Teyssier+2013} and \citet{Governato+2015}. We have further measured 
the median proper radius (with respect to the galaxy center) of all stellar particles inside $0.02$\rh\
at 5 Gyr age ($z=1.2$), then traced these particles to $z=0$, and measured their median proper radius again. 
For the WDM run, the median radius increased from 0.15 to 1.44 kpc; 
that is, by nearly a factor of 10. The same measures for the CDM run, show an increase from 0.32 to 0.70 kpc;
that is, only by a factor of about two. Therefore, {\it the stellar particle distribution in the inner regions 
of the galaxy in the WDM run dramatically expands with respect to the modest expansion seen in the CDM run. }

\begin{figure}
\plotdelgado{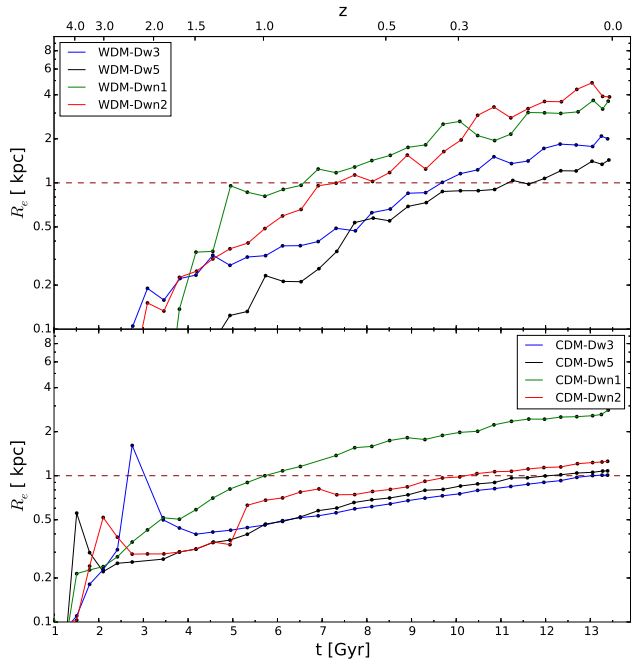}
\caption{Evolution of the stellar half-mass radius, \re, for the four WDM (upper panel) and CDM (bottom panel) runs.}
\label{re-evol}
\end{figure}

In Fig. \ref{re-evol}, the evolution of the half-mass radius of our four dwarfs is plotted, both for the WDM (upper panel)
and the CDM (lower panel) runs. The \re\ growth of the WDM dwarfs is much more pronounced than the growth of the
corresponding CDM dwarfs. A growth of \re\ is expected due to the inside-out mass buildup of galaxies
in the hierarchical clustering scenario \citep[see, e.g.,][]{Firmani+2009}. For example, according to the
models presented in that paper, the \re\ of low-mass galaxies grows on average by factors of $\approx 2-2.5$
from $z=1.5$ to 0. This factor is $\approx 2.5- 5$ for our CDM dwarf simulations, and it is likely due to both the inside-out
mass assembly of the galaxies and a slight inner expansion. In the case of the WDM simulations, this factor
is much larger, it is in between $\approx 7$ and 21, evidencing that besides the inside-out galaxy assembling, \re\
grows and the inner stellar surface density decreases due to an expansion of the stellar particle distribution.
According to our simulations, while in the CDM case (no shallow core formation) the central stellar surface density 
increases as the galaxy grows inside-out, in the WDM case (shallow core formation) this increase is lower,
and at later epochs, it even decreases; for Dw5, Dwn1, and Dwn2, from $z\sim 1$ to 0, the central surface
density decreases by factors of $3-20$. 

It is quite interesting that the dynamical mechanisms able to expand the inner dark matter mass distribution, producing
a shallow core, also have consequences for the stellar mass distribution: the stars that formed early in 
the centers of the dwarfs migrated to orbits with larger radii, and then the stellar age and metallicity gradients become flatter 
than in the absence of the expansion. Besides, the radial stellar surface density profiles become ``flatter'',
with a nearly exponential distribution and a lower amplitude in the center (see Fig. 4 in Paper I), 
as first discussed in \citet[][]{Read+2005}. 

Our results show that the dynamical heating mechanisms can affect the inner stellar 
particle spatial distribution even more  
than that of the dark matter one (see upper panel of Fig. \ref{fluctuations}).
This might be because the velocity distributions of the stellar particles are more radially anisotropic than those
of the dark matter particles, and the former become more easily unbound than the latter as a result of rapid changes 
in the local potential \citep[][]{Read+2005}. Therefore, {\it along with the formation of a shallow halo core, important
signatures are imprinted in the observed dwarf galaxies, such as in the color, age, and metallicity gradients, as well as 
the surface density (brightness) profiles. }

\section{Summary and discussion}
\label{summary}

We have analyzed the evolution of  WDM simulations (and their corresponding CDM counterparts)
of central dwarf galaxies formed in distinct halos of $\mh=2-3\times 10^{10}$ \msunh\  at $z=0$. 
These zoom in simulations, which were presented in Paper I, 
were performed with the ART code using the subgrid physics
first employed in \citet{Colin+2010}. Our aim is to explore and compare the effects of baryons on the
inner halo structure in the \lwdm\ and \lcdm\ cosmologies at a given subgrid physics. 
The masses of the systems in the WDM case are close to the filtering mass, which corresponds 
to a thermal particle of 1.2 keV. The main results regarding the evolution of the inner structure
and dynamics of the halos and stellar galaxies are as follows.

\begin{itemize}

\item At $z=0$, the inner halo density profiles of the WDM dwarfs have shallow inner slopes \alp2\ 
with values between $-0.25$ and $-0.66$. This flattening of the inner profiles is not
observed in the corresponding CDM simulations. It is confirmed that in the case of the DM-only simulations,
both the CDM and WDM halos are reasonably well described by the NFW profile but the 
latter are significantly less concentrated than the former by factors of 1.5--1.9. 

\item The halo 3D velocity dispersion profiles of the WDM simulations including baryons tend to flatten
in the inner regions, evidencing a dynamical heating of the dark matter particles at these regions 
with a consequent expansion. 
Indeed, the pseudo phase-space halo density profile, $Q(r)=\rho/<\sigma_{\rm 3D}^2>^{3/2}$, of 
these runs flattens systematically in the center with time, deviating from the initial power law
with a slope $\approx -1.9$. On the other hand, for the CDM counterpart runs, $Q(r)$ hardly 
changes with time. It is described by a power law with a slope $\approx -1.9$ (typical of a NFW configuration), 
both in the dark-matter-only and the hydro simulations.

\item The inner slope \alp2\ of the WDM dwarfs in the hydro simulations at $z= 2.5$ have values in
between $-1.1$ and $-1.3$, in epochs when the galaxies just started to form (the \ms-to-\mh\ ratios are
much lower than 0.1\%).  Then, the slopes slightly increase on average and at some point, in periods 2--5 Gyr 
long, they strongly increase, becoming the halo cores shallow, and remaining as such during the 
last $3-5$ Gyr. Their CDM counterparts at $z = 2.5$, when the SF activity is high, show halo
inner slopes \alp2\ in between $-0.7$ and $-1.1$, but then decrease and finally remain 
roughly constant at values of $-1.2\div -1.5$. 

\item At any epoch, when \alp2\ is larger than $-0.5$ (shallow core), the integral SF efficiency, \ms/\mh,
has values in the range $0.25-1$\%, though having such values does not imply a shallow core; therefore,
a necessary but not sufficient condition for shallow core formation is to have \ms/\mh\ values in this range.
On the other hand, when \ms/\mh\ is lower than 0.1\%, \alp2\ is steeper than $-0.9$.

\item The case study of evolution presented in detail for the Dw5 run, both in the WDM and CDM cosmologies,
shows that the behavior of the inner slope \alp2\ with time does not correlate significantly
with the SFR history, the gas outflow rate from the galaxy or the gas outflow-to-inflow rate.
In particular, during the epochs when \alp2\ gets significantly shallower (ages from 5 to 9 Gyr),
the average SFR and the outflow rates even decrease. However, the SF and gas outflow rates are very 
bursty at all epochs, significantly more than in the corresponding CDM run.

The episodic gas outflows due to the bursty SF history produce strong fluctuations of the gas
content inside a sphere of 500 pc around the dark matter halo center, and in many cases
these fluctuations are associated with bulk gas motions. The fluctuations can have amplitudes as
large as 100--200 at the epochs when efficient inner halo profile flattening is happening in the 
analyzed Dw5 WDM run. Although the gas mass inside this sphere does not contribute more than
$\sim 1/3$ of the total mass, the fluctuations are so strong and quick that they likely produce 
non-negligible fluctuations in the inner gravitational potential. The quick fluctuations 
in the potential and the bulk gas motions are conditions proposed to dynamically heat the dark 
matter particle orbits and produce core expansion. At later times, when \alp2\ 
remains constant on average, the gas mass fluctuations inside 500 pc become
lower than at early epochs. For the Dw5 CDM run, the gas mass fluctuations are much less
intense than in the WDM run; the highest fluctuations happen at the earliest epochs when
a slight flattening of the inner halo is seen. 

\item The possibility of angular momentum transference from the stellar particles to the dark matter ones has also been
explored by measuring the specific angular momentum of dark matter and stellar particles inside 1 kpc
for the WDM Dw3 and Dw5 galaxies. While $j_{\rm DM}$ remains almost the same with small
fluctuations, $j_s$ evolves in different ways for each case and with large fluctuations, without evidencing
a process by which angular momentum is transferred from stellar to dark matter particles.

\item Not only the inner dark matter halo but also the
inner parts of the stellar galaxy expand in the WDM hydro simulations. 
As a consequence, the half-mass radii, \re, of the dwarfs strongly 
grow with time, by factors of $7-21$ since $z=1.5$. The growth is much weaker for the CDM 
counterpart simulations, by factors $2.5-5$, which is mainly due to the inside-out growth of the 
galaxies. While the central surface brightness of these dwarfs remains roughly
constant since $z\sim1$, for the WDM dwarfs, it decreases by factors of $3-20$. The
dynamical mechanisms that expand the halo cores also expand the stellar galaxies,
probably even more efficiently.

\end{itemize}

The main conclusion from our numerical simulations is that the flattening of the inner halo
density profile happens due to the strong fluctuations of the amount of gas in the inner
regions which induce quick variations in the central gravitational potential and
bulk motions of the gas with respect to the halo center; both seem to be efficient
mechanisms of kinetic energy transfer to the dark matter particles 
\citep[][]{Mashchenko+2006,Pontzen+2012}. {\it These mechanisms also
produce a strong expansion of the stellar particle spatial distribution and a late
decrease of the central stellar surface density. }Then, (1) the stellar age and metallicity
gradients of a present-day dwarf are expected to be flat or even positive despite 
the fact that the galaxy was assembled inside-out, and (2) the stellar surface density of the dwarf results 
in a ``flatter'', exponential-like profile , and with a  lower amplitude in the center than in the absence
of the dynamical expansion mechanism. 

For the subgrid physics used here in the hydro simulations, a significant inner expansion of the dark matter 
and stellar particle distributions occurs only in our dwarfs in the WDM cosmology, not in 
the CDM one. The WDM halos of masses close to \mhm\ assemble later and are significantly less 
concentrated than their CDM counterparts; thus, the gravitational potential is less deep in 
the former than in the latter.  Besides--as a consequence of what we have just said--the SF history in the 
WDM simulations is delayed with respect to and much burstier than in the CDM case (see also Paper I). Therefore, 
{\it late halo assembly, low halo concentration, and a late, sustained and bursty SF history seem to be the 
conditions required for shallow core formation in low-mass halos} \citep[see also][]{Onorbe+2015}.  
These conditions were not fulfilled in our four zoom-in \lcdm\ simulations of systems that end with
 virial masses $\mh\approx 2-4\times 10^{10}$ \msun. 
 
 However, there are (rare) cases when the mass assembly history of halos of these masses
 in the \lcdm\ cosmology are more extended in time, implying lower halo concentrations and later 
 and more episodic SF histories than the average. For example, this was the case 
 of runs Dw6 and Dw7 in \citet{Gonzalez-Samaniego+2014}, who explored  the evolution and properties of 
 simulated dwarfs formed in halos of $2-3\times 10^{10}$ \msun\ for a broad range of mass 
 assembly histories. We measured the \alp2\ inner slope of the halos in runs Dw6 and Dw7 and
 found that they are the shallowest among the seven isolated systems presented in that paper, with
 values at $z=0$ of $\alp2=-0.52$ and $-0.88$, respectively. The former halo has values of \alp2\
 around $-0.5\div -0.6$ since $z\sim2$, while the latter halo shows an increase in \alp2\ from $\approx -1.3$ to
 $-0.88$ since $z\sim 0.7$ to 0 (this halo suffered a major merger at $z\approx 1.8$, after which
 \alp2\ decreased from $\alp2\approx -1$ to $-1.3$ up to $z\sim 0.7$, when the flattening starts). 
In conclusion, our simulations show that the inner density profile of dwarf galaxy halos may significantly flatten
in the \lcdm\ cosmology only in the extreme cases of late and extended in time mass aggregation
histories, which imply low-concentration halos and strongly episodic and sustained SF histories. 

\subsection{WDM vs. CDM}

Our numerical experiments show that the formation of shallow cores by the effects of baryons
in halos of present-day masses $\mh\sim 10^{10}$ \msun\ is much easier in the WDM cosmology
(for \mhm\ close to $\sim 10^{10}$ \msun), than in the CDM one. For the filtering mass \mhm\ used in our
simulations, the corresponding thermal relic WDM particle mass is \mwdm=1.2 keV (for a non-thermal
particle, such as the non-resonantly produced sterile neutrino, the respective particle mass in our
cosmology is 10.7 keV according to \citealp{Viel+2005}).  As discussed in Paper I, the dwarf galaxies
simulated in this cosmology are in better agreement with observations than their CDM counterparts, 
except for their too large half-mass radii. However, by comparing the results of WDM hydrodynamic simulations
in the quasi-linear regime with the Ly-$\alpha$ flux power spectrum of high-redshift quasars,
it turns out that $\mwdm \simgreat 3$ keV \citep{Schneider+2013,Schneider+2014}. 
Note that these comparisons are not free of uncertainties and limitations (see, e.g., \citealp{deVega+2014, Garzilli+2015}). 
For example, the latter authors find from a re-analysis of high-redshift and high-resolution Lyman-$\alpha$ 
forest spectra that thermal relics of mass $\mwdm=2-3$ keV can provide as equally good fits as CDM depending 
on the thermal history of the intergalactic medium.

In Paper I we have also presented simulations for a WDM cosmology with $\mwdm=3$ keV.
For this particle mass $\mhm=1.4\times 10^9$ \msun\ ($h=0.7$) and the present-day virial masses of the
simulated systems were similar to those presented here ($2-4\times 10^{10}$ \msun); that is, these
systems are 20--30 times larger than \mhm. The main properties and evolutionary trends of these
runs were close to those of their CDM counterparts. We measure here the inner halo \alp2\ 
slope for these WDM 3.0 keV runs and find that in two cases (Dw3 and Dw4) they are only slightly shallower than those
of their CDM counterparts, with values of $\alp2\approx -1.2$ at $z=0$. However, in the other two cases (Dw5 and Dw7), 
the inner slopes are significantly  shallower  than in their corresponding CDM simulations, with values 
of $\approx -0.8$ and $-0.5$, respectively. When comparing the evolution of \alp2\ in the WDM 1.2 and 3.0 keV
simulations for the Dw5 dwarf, we see a similar trend: \alp2\ becomes shallow with time after $\sim 5$ Gyr,
though at a slower rate in the latter case. 

Thus, in a WDM cosmology with $\mhm\sim10^9$ \msun\ ($\mwdm\simgreat 3$ keV), it is possible that a significant fraction 
of the present-day halos of masses $\sim 2-4\times 10^{10}$ \msun\ may have formed shallow cores. For less massive halos, 
the formation of shallow cores could be more common. Following the results presented in this paper, for halo masses close to 
\mhm, we would expect that all the halos would form shallow cores. However, for lower mass halos, the SF reduces dramatically.
We have run hydro simulations for a dwarf that ends at $z=0$ with 
$\mh=4.2\times 10^9$ \msun\ (for lower masses halos are resolved with too few particles) both in this cosmology and in the 
CDM one.  In fact, the inner halo density profile does not flatten in both cases; the measured inner slope \alp2\ at $z=0$
is $-1.3$ and $-0.95$, respectively; that is, in the WDM case it is even steeper than in the CDM one. We believe 
this is so because the SF in these very low-mass halos is strongly suppressed, even more in the WDM run than in the
CDM one. The WDM dwarf ends with a stellar mass of $1.52\times 10^{6}$ \msun, 14 times lower than the CDM one. 
Therefore, the effects of baryons are negligible, at least for the resolution that attain these simulations,
which is similar to the one achieved by the dwarfs in the 1.2 keV WDM cosmology.

From the results found here and in Paper I, we conclude that in a $\sim 3$ kev WDM cosmology, the dwarf galaxies formed in halos of masses $\sim 10-20$ times the corresponding filtering mass, $\mhm\sim 10^9$ \msun, have properties only slightly different from those of their CDM counterparts, but in the former case the halos are more prone to form shallow cores than in the latter one. For masses lower than $\mh\sim 2-4\times 10^{10}$ \msun\ ($\ms\sim 3-5\times 10^8$ \msun), most of the halos are expected to have shallow cores due to the effects of baryons. However, for masses lower than $\sim 5\mhm$ ($\mh\simless 5\times 10^{9}$ \msun, $\ms\lesssim 2\times 10^6$ \msun), the SF is strongly suppressed so  that the effects of baryons seem to be negligible in the inner dark matter structure\citep[][but see also \citealt{Onorbe+2015}]{Governato+2012,diCintio+2014}.
In general, in the WDM cosmology a larger diversity of inner halo density profiles than in the CDM cosmology is expected 
because for masses much larger than \mhm, the density profiles are very similar in both cosmologies, while for lower
masses, in an increasing fraction of cases, the inner halo profiles flatten in different degrees in the WDM cosmology, widening
the range of inner circular velocities with respect to the CDM case. The observations seem to present a significantly
larger diversity of inner circular velocities than those measured in a \lcdm\ hydro cosmological box simulation 
\citep[the EAGLE and LG simulations;][]{Oman+2015}.  In order to attain a full statistical description of the low-mass galaxy/halo 
population in the WDM cosmology, hydro simulations in cosmological boxes are necessary.

\section*{Acknowledgements}            
We are grateful to the anonymous Referee for his/her constructive suggestions that enriched the manuscript.
AG acknowledges a Postdoctoral fellowship provided by the CONACyT grant (Ciencia B\'asica) 180125. 
VA and PC acknowledge CONACyT grant (Ciencia B\'asica) 167332.

\bibliographystyle{apj}
\bibliography{references_WDM}
\label{lastpage}
\end{document}